\documentclass{sig-alternate-2013}

\usepackage{amsmath}
\usepackage{xcolor}
\usepackage{graphicx}
\usepackage{subcaption}
\usepackage[labelfont=bf,font=bf]{caption}
\usepackage{float}
\usepackage[hyphens]{url}

\newtheorem{theorem}{Theorem}[section]

\newtheorem{lemma}[theorem]{Lemma}

\newtheorem{definition}[theorem]{Definition}

\newcommand{\TL}[1]{
\mathrm{TL}_{#1}
}

\newcommand{\precision}[1]{
\mathrm{Precision}_{#1}
}

\newcommand{\TLU}[1]{
\mathrm{TLU}_{#1}
}

\newcommand{\squishlist}{
 \begin{list}{$\bullet$}
  { \setlength{\itemsep}{0pt}
     \setlength{\parsep}{3pt}
     \setlength{\topsep}{3pt}
     \setlength{\partopsep}{0pt}
     \setlength{\leftmargin}{1.5em}
     \setlength{\labelwidth}{1em}
     \setlength{\labelsep}{0.5em} } }
\newcommand{\squishend}{
  \end{list}  }

\graphicspath{{Images/}}

\permission{\copyright 2017 International World Wide Web Conference Committee \\ (IW3C2), published under Creative Commons CC BY 4.0 License.}
\conferenceinfo{WWW 2017,}{April 3--7, 2017, Perth, Australia.}
\copyrightetc{ACM \the\acmcopyr}
\crdata{978-1-4503-4913-0/17/04. \\
http://dx.doi.org/10.1145/3038912.3052647 \\
\includegraphics{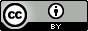}
}
\begin{document}


\title{Competition and Selection Among Conventions}

\numberofauthors{4}
\author{
\alignauthor
Rahmtin Rotabi\\
       \affaddr{Cornell University}\\
       \email{rahmtin@cs.cornell.edu}
\alignauthor
Krishna Kamath\\
       \affaddr{Twitter Inc.}\\
       \email{kkamath@twitter.com}
\alignauthor
Jon Kleinberg\\
       \affaddr{Cornell University}\\
       \email{kleinber@cs.cornell.edu}
\and  
\alignauthor
Aneesh Sharma\\
       \affaddr{Twitter Inc.}\\
       \email{aneesh@twitter.com}
}

\title{Cascades: A View from Audience}

%
\maketitle

\begin{abstract}
  Cascades on social and information networks have been a tremendously
  popular subject of study in the past decade, and there is a
  considerable literature on phenomena such as diffusion mechanisms,
  virality, cascade prediction, and peer network effects.  Against the
  backdrop of this research, a basic question has received
  comparatively little attention: how desirable are cascades on a
  social media platform from the point of view of users?  While
  versions of this question have been considered from the perspective
  of the {\em producers} of cascades, any answer to this question must
  also take into account the effect of cascades on their audience ---
  the viewers of the cascade who do not directly participate in
  generating the content that launched it.  In this work, we seek to
  fill this gap by providing a consumer perspective of information
  cascades.

  Users on social and information networks play the dual role of
  producers and consumers, and our work focuses on how users perceive
  cascades as consumers. Starting from this perspective, we perform an
  empirical study of the interaction of Twitter users with retweet
  cascades. We measure how often users observe retweets in their home
  timeline, and observe a phenomenon that we term the {\em Impressions
    Paradox}: the share of impressions for cascades of size $k$ decays
  much more slowly than frequency of cascades of size $k$. Thus, the
  audience for cascades can be quite large even for rare large
  cascades. We also measure audience engagement with retweet cascades
  in comparison to non-retweeted or organic content. Our results show
  that cascades often rival or exceed organic content in engagement
  received per impression. This result is perhaps surprising in that
  consumers didn't opt in to see tweets from these
  authors. Furthermore, although cascading content is widely popular,
  one would expect it to eventually reach parts of the audience that
  may not be interested in the content. Motivated by the tension in
  these empirical findings, we posit a simple theoretical model that
  focuses on the effect of cascades on the audience (rather than the
  cascade producers). Our results on this model highlight the balance
  between retweeting as a high-quality content selection mechanism and
  the role of network users in filtering irrelevant content. In
  particular, the results suggest that together these two effects
  enable the audience to consume a high quality stream of content in
  the presence of cascades.
\end{abstract}

%
%
%

%
%

%
%



\section{Introduction}

Users on modern social and information networks play dual roles as
content producers and consumers: content they produce is seen by their
friends or followers, and content they see (or consume) on the network
is produced by users they are friends with or following. In addition
to producing their own content, these networks also provide users with
low-friction content-producing mechanisms. Users can switch from being
consumers to content producers with a single click as they can share
or retweet content that they want to communicate to their followers. In
some cases, just consumption activity (such as ``liking'') is akin to
content production from the user in terms of what their
followers/friends observe.

Having a low barrier for content production is clearly important in
activating the information-sharing aspects of social and information
networks, but some of these mechanisms could be viewed as existing in
tension with a basic contract of these networks.
A key premise of a social or information
network is that users opt in to connect to friends or users that they
are explicitly interested in hearing from. But in the presence of
sharing mechanisms, cascades originate on the network and hence a user
could often see content from users they did not opt in to see content
from. It is conceivable that cascades could overwhelm a user's home
timeline, rendering the network significantly less
useful to the user. Indeed, when
retweets were first introduced on Twitter, users expressed many such
concerns~\cite{RetweetSucks09}.

A key question then is: what effect do cascades have on consumption
behavior? A pithy answer is provided by the existence of
networks with hundreds of millions of active users; this at least suggests
that the effect of cascades is not as negative as users feared it to be. Our
work aims to quantify this effect, and provide some insight into why
this might be the case.

One aspect of user consumption behavior is deeply intertwined with
production in that production of content (via re-sharing) is also
simultaneously consumption behavior. Production has been widely
studied in the literature under the topics of information propagation
and diffusion of content. We note however that this is only a part of
consumption and some basic characteristics of consumption behavior
have not been addressed to the best of our knowledge. For instance,
although virality of content on Twitter has been extensively
discussed~\cite{goel-structural-virality,Wu2011}, we do not understand
the view of virality from a consumer perspective: what fraction of
tweets consumed by consumers on Twitter are viral? Do users engage
with these more than with non-viral content in their home timeline? We
emphasize that the consumer view could be quite different from the
producer perspective for virality: even though a small fraction of
tweets ``go viral'', a large fraction of the consumer experience on
Twitter could still be shaped by viral content.  This is because when
we think about the population of all {\em views} of tweets, we're
sampling tweets in proportion to their popularity, and this sampling
based on size leads to effects where a small number of items
(extremely popular tweets in this case) can make up a large fraction
of the sample \cite{arratia-size-bias}.

We examine the above questions through empirical analysis of user
behavior on Twitter.  Each time a user views a tweet --- referred to
as an {\em impression} of the tweet --- they can choose to engage with
it through several means, including clicking on it, liking it, or
retweeting it.  Given observations of what tweets a user sees in their
home timeline via tweet impression logs, as well as tweet engagement
and sharing activity, we can piece together a consumer view of
cascades on Twitter. Through this analysis, we observe that retweet
cascades indeed occupy a substantial fraction (roughly a quarter)
of a typical
user's timeline, and 1 out of 3 impressions in the dataset we analyze
are due to cascades. Thus, cascades have a substantial impact on the
user experience at Twitter given their prevalence in users' home timelines.
This impact is arising despite the fact that extremely few tweets
generate large cascades; the point is that for a producer of content,
it is very rare to see your tweet become viral, but for a consumer of
content, much of your time is spent looking at viral content.
We term this dichotomy the {\em Impressions Paradox}; it is a
counter-intuitive contrast in two ways of looking at the same
population, in the spirit of similar phenomena that arise because
of sampling biased by size.

In light of our previous discussion, we note that this wide
prevalence of retweets does indeed impose upon users content that they
did not opt in to see. It is natural to wonder whether users respond
negatively to this imposition on their home timeline, which they have
carefully constructed through their choice of users to follow.

Analyzing user engagement with cascades provides a way to answer this
question. In particular, we compare user engagement probabilities
(retweeting, liking and clicking) on retweeted content versus organic content
(directly produced by a person the user follows). Our main
finding here is that retweeted content rivals or exceeds the organic
content in engagement.
It is useful to consider this fact in the context of
user fears of irrelevant content showing
up in their timeline (even if it might be high quality; the best tweet
on politics may be uninteresting to a user not interested in politics).
Viewed in this light, our finding is perhaps quite unexpected. On the other
hand, this result is exactly what one might expect if we think of retweets
as a high quality tweet selection mechanism --- users might only
engage with the best tweets, so it is unsurprising that the best
tweets get high engagement.
Note however, that popular tweets also get viewed by many users, resulting
in a very high number of impressions.
Thus, even with an assumption that popular tweets have high quality,
it seems unclear why they should get high engagement {\em per
impression} as their growth in audience size might completely
outpace the set of interested users.

In order to understand this effect quantitatively, we propose and
analyze a simple theoretical model of retweeting behavior that teases
apart these two effects. Our model is novel in that it inverts the
traditional view of cascades as a tree being rooted at the author, to
a tree that is centered at an arbitrary user --- a member of
the audience for cascades --- who receives a mix of organic
tweets and retweets. This model helps us quantify two metrics for a
user's home timeline: precision (seeing content that is relevant or
topical for users) and quality (highly engaging content for a
topic). Intuitively, users would like to have a high precision and
high quality home timeline where most of the content is relevant and
highly engaging. In the presence of retweets, it seems unclear a
priori whether the content will still be relevant, and further how
would one quantify changes in tweet quality. Our analytical and
simulation results show that it is indeed possible for users to have
the best of both worlds by seeing high quality and relevant retweets
in their timeline. Furthermore, this model also helps us understand
the {\em value} of retweets by quantifying a counterfactual world
where retweets would not exist.

\section{Related work}
There has been extensive work on on-line information diffusion.
This has included studies of news
\cite{adamic2005political,Berger:JournalOfMarketingResearch:2012,Bakshy:Science:2015},
recommendations \cite{leskovec-ec06j},
quotes \cite{DanescuNiculescuMizil:ProceedingsOfTheAcl:2012},
hashtags on Twitter
\cite{Romero:ProceedingsOfThe20ThInternationalConferenceOn:2011,Tsur:ProceedingsOfWsdm:2012,Romero:ProceedingsOfIcwsm:2013,Lin:Icwsm:2013,Maity:ProceedingsOfCscw:2016},  information flow on Twitter \cite{Wu2011}
and memes on Facebook \cite{DAF13,cheng-www14}.
Past work has also investigated methodological issues including
definitions of virality \cite{GAHW15},
the problem of prediction \cite{cheng-www14},
the trade-off between precision and recall in cascading content \cite{BGMS13},
and the role of mathematical epidemic models \cite{GMSZ15}.


In addition, it has been shown that only a very small fraction of
cascades become viral \cite{GAHW15} but the ones that do become viral
cover a large/diverse set of users. In other words, if you are the
source of a cascade you have a low chance of creating a viral cascade
but, once we switch to the consumer's point of view we observe that a
large fraction of a user's timeline is made up of these diffusing
pieces of content. Another related theme on this work has been the
observation that a small number of ``elite'' users produce a
substantial fraction of original content on Twitter~\cite{Wu2011}. As
with other studies, this one also focused on active cascade
participants, and our work is differentiated by the focus on cascade
audience.

The primary focus of the body of prior work on cascades has been either
on the source of the content or on the structural properties of
the cascades themselves. In this work,
we study the effect of different properties of the cascade tree, and
the underlying follower graph, on the experience of the consumers of cascades.
In particular we first show that although most tweets do not
get re-shared but they are a significant fraction of the content an
average user reads. We also find that consumers prefer either very
popular content or personalized content coming from users they opted
to follow. Then we look at each consumer as an individual and show
that different consumers might show different behavior but a single
user is consistent on the type of content they like over several
days. Finally, we complete our argument with a simple model that
captures how the re-share mechanism features enhance the experience of
consumers.

\section{Empirical Analysis on Twitter}
In order to analyze audience behavior with cascades, we undertook an
empirical investigation on Twitter.  A user $u$ on Twitter typically
spends the majority of their time on their personalized home page,
called the home timeline, which primarily consists of a collection of
{\em Tweets} from a set of users $F(u)$ that $u$ chooses to
follow. Since most content is consumed on the home timeline, we focus
our analysis on user behavior in the home timeline. Further, to keep
our analysis most interpretable, we ignore some products that rank the
Twitter home timeline, such as ``While you were away''
\cite{TwitterWhileAway}, and focus exclusively on impressions of
unranked tweets. These tweets are presented in a reverse-chronological
fashion, and hence they allow us to study a version of the question
that is independent of the ranking process (since ranking can have a
large implication for the visibility and hence effect of cascades; see
e.g. Facebook's work on this issue \cite{FacebookBlog}).

We measure both views (or impressions) of tweets as well as user
engagements with the tweets in our analysis. We define these terms in
the sections below, but the goal of the analysis is to provide insight
into the impact of cascades on consumer experience via impressions,
and gauge their reception of this content via engagement. The dataset
for this analysis was collected from Twitter logs during a $16$ day
period during summer 2016.  Because of user privacy, we conduct all
the analysis in a user-anonymized fashion, and present results from
aggregate analysis. Note that for some of the plots we use a relative
scale for the $y$-axis to anonymize actual values, as a relative
comparison of values is the main goal for these plots.

\subsection{Cascade Views}
The first step in our empirical investigation is to understand whether
cascades constitute a significant fraction of overall audience
attention. Perhaps the simplest metric for measuring this is to
understand the raw volume share in a user's home timelines. But before
we proceed with that, we need to define what we mean by a tweet cascade and
what constitutes a ``view'' of a tweet.

\begin{figure}[hbtp]
\begin{subfigure}[t]{.23\textwidth}
\centering
\includegraphics[width=\linewidth]{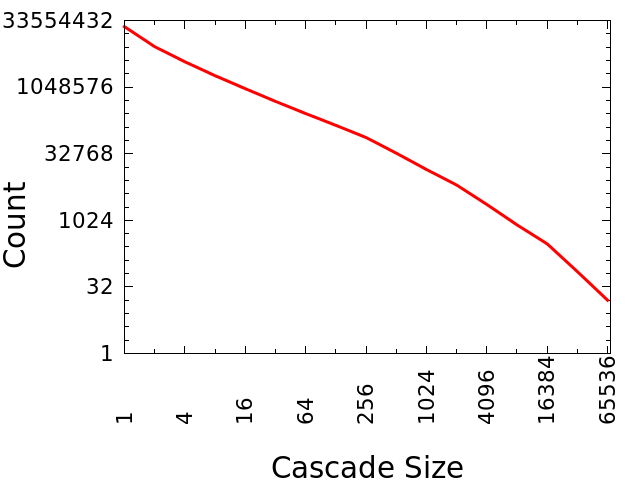}
 \caption{}\label{fig: CascadeDistribution}
\end{subfigure}
\begin{subfigure}[t]{.23\textwidth}
\centering
\includegraphics[width=\linewidth]{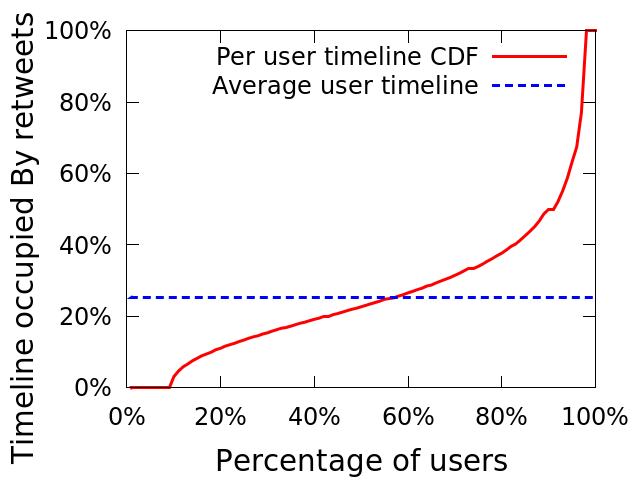}
\caption{}\label{fig:retweet_impressions}
\end{subfigure}
\caption{(a) The number of cascades in a log-log plot bucketed using the $\lfloor log_2 \rfloor$ function
(b) The distribution of the fraction of home timeline
      impressions constituted by retweets, over all Twitter users. The
    horizontal line represents the average value.
    }
\end{figure}

In this work any tweet that is retweeted at least once is called a cascade
as it has one retweeter (or adopter) other than the original
tweeter. As has been noted in prior work, most cascades are shallow
(star-like) and only a rare few go on to be ``viral''. We also refer
to a {\em cascade size}, which is the number of eventual retweeters
(or adopters) that the tweet gathers, as measured after a few days of
the original tweet. In order for our analysis to be valid
the dataset must be large enough to catch cascades from a
spectrum of sizes. We confirm this by visually plotting the cascade
size distribution in Figure~\ref{fig: CascadeDistribution}, which
shows the average number of cascades with size between $[2^k,2^{k+1})$
over the 16 days. Clearly, the data has enough cascades in each
bucket even for a single day.

Next, we define a tweet view or an {\em impression} on the home
timeline. The ideal measurement is to check that the user really
``saw'' the tweet, but in absence of that, we just measure whether
 the tweet stayed on the user's mobile screen
for a large enough time. This filters out a variety of behaviors, and
among them the common pattern of scrolling quickly through the home
timeline, where the user just glances at a large number of tweets.

With these definitions we turn our attention to studying how much
audience attention is commanded by cascades. Perhaps the most basic
measurement to make is to measure what fraction of a users' home
timeline impressions came from cascades that did not originate in the
user's direct neighborhood. We find that 68\% of all home timeline
tweet impressions are from users' direct followings, and the remainder
32\% come from cascades that originate from outside of a users' direct
neighborhood. A different view of this overall statistic comes from
looking at this from each individual user's perspective,
through which we
can measure what fraction of a user's timeline impressions come from
retweets. The distribution of this quantity is shown in
Figure~\ref{fig:retweet_impressions}, from which it is evident that
for half of all users, approximately a quarter of their timeline
consists of retweets. An additional dimension of tweet impressions
coming from cascades is that these tweets bring in a fair bit of
author diversity: 55\% of unique authors who appear in a user's
timeline are from outside the user's direct followings.

Given that nearly a quarter of a user's home timeline consists
of cascades, it is natural to ask how these cascades reach the
user. To provide insight into this question, we look at how far away
the cascade originated, and how long it took to get to the
user. For the former, we measure the network {\em distance} (shortest
directed path) from the user to the cascade originator (the author of
the root tweet in the cascade). Table~\ref{table: ImpressionsOnDistanceAndHopCount}
shows the percentage of tweet impressions
that occur for each distance (out of all impressions), and from the
data it is clearly visible that almost all impressions come from
within distance $2$ in the graph.

\begin{table}
\centering
\caption{The percentage of tweets on a timeline based on their distance and hop-count to the receiver}
\begin{tabular}{|c|c|c|c|c|}
\hline
Distance & Impressions & &Hop count & Impressions \\ \hline
1	& $68.86\%$ & 	& 1&$66.70\%$  \\ \hline
2	& $30.53\%$ &	& 2& $27.48\%$	\\ \hline
3	& $0.59\%$ & 	&3 & $3.66\%$		\\ \hline
4	& $10^{-3}\%$ & 	 &	4& $1.09\%$		\\ \hline
5   & $10^{-4}\%$ & 		 &	5& $0.45\%$ \\ \hline
6	& $4 \times 10^{-5}\%$ &    &6	& $0.23\%$    \\ \hline
7	& $6 \times 10^{-6}\%$ &   &	7& $0.13\%$      \\ \hline

\end{tabular}
\label{table: ImpressionsOnDistanceAndHopCount}
\end{table}

To understand the path the cascade took to reach a user, we
reconstruct the cascade tree\footnote{The cascade is a directed
  acyclic graph from a user's perspective, but we can think of it as a
  tree by picking the first incoming edge for each node by time ---
  that is also a close approximation to how Twitter treats retweets in
  practice.} and compute the number of hops on the tree
that are between the user and the root of the tree;
we refer to this quantity as the {\em hop-count}.
The distribution of impressions w.r.t the hop-count is shown in
Table~\ref{table: ImpressionsOnDistanceAndHopCount}. As is the case with distance, almost all
impressions occur on hop counts 1 and 2. This data is all in agreement
with prior work that has also commented on the vast majority of
cascades being very shallow in terms of
hop-count~\cite{GAHW15}. However, we do note that in contrast with
distance, impressions for larger hop-counts don't quickly die down to
zero. An obvious hypothesis for this is the possibility that some
large cascades survive for a long time and hence reach users via all
kinds of paths. This leads to the question of the impact of these
large cascades from the perspective of impressions.

\begin{figure}[hbtp]
\begin{subfigure}[t]{.23\textwidth}
\centering
\includegraphics[width=\linewidth]{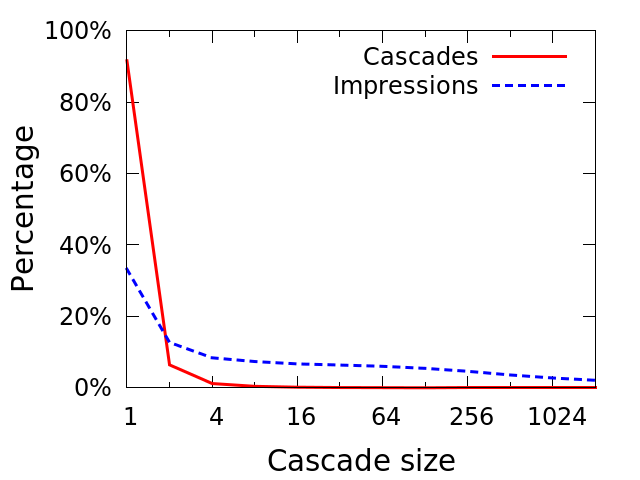}
 \caption{}\label{fig:ImpressionParadox}
\end{subfigure}
\begin{subfigure}[t]{.23\textwidth}
\centering
\includegraphics[width=\linewidth]{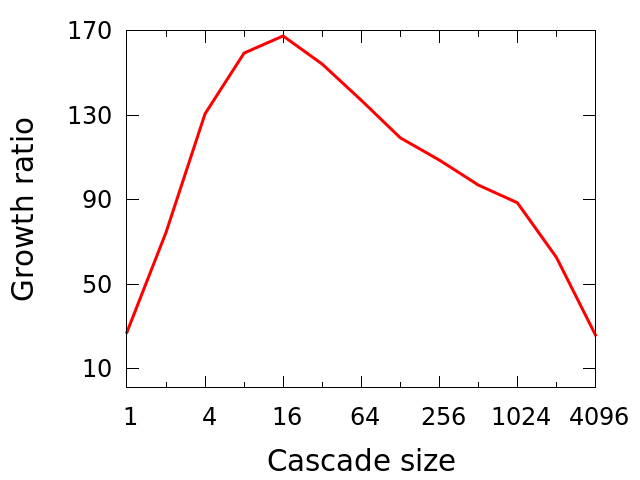}
\caption{}\label{fig: GrowthRatio}
\end{subfigure}

\caption{(a) Illustrating the {\em Impressions Paradox}: share of
     impressions for cascades of size $k$ decays much more slowly
      than frequency of cascades of size $k$. Note that x-axis is log
      scale. (b) Cascade growth ratio is the ratio between number of
      impressions generated by these cascades to number of tweets
      generated by these cascades.}
\end{figure}

Recall that 1 out of 3 impressions arise from cascades, and in light of
the previous discussion, we would like to understand {\em which} kinds
of cascades are contributing to these impressions. It is useful to
remember that large cascades are quite rare on Twitter, as illustrated
in Figure~\ref{fig:ImpressionParadox}: on a log-percentage plot, the
fraction of tweets that get more than 8 retweets is less than
1\%. However, since some cascades survive for a long time, it is
natural to ask what is the share of impressions generated by these
cascades. The share of impressions is also shown on the same
Figure~\ref{fig:ImpressionParadox}, and this presents a stark contrast
from the probability of tweets generating a large cascade --- even
though 91\% of tweets have a cascade size of 1, these generate only
33\% of impressions coming from retweets, with the large cascades
contributing a substantial fraction of impressions coming from
retweets. We term this the {\em Impressions Paradox}: the share of
impressions for cascades of size $k$ decays much more slowly than
frequency of cascades of size $k$.


We also note that for all cascades of a given size, one can compute a
{\em cascade growth} metric: the ratio between number of impressions
generated by these cascades to number of tweets generated by these
cascades.
Intuitively, one would expect this ratio to be high for
small cascades since almost every retweet brings in a large set of new
audience members who haven't seen it by other means,
while for larger cascades the gain in new audience per retweet might be
lower since the presence of triangles means that new retweets may
eventually reach people who have already seen it by other means.
In fact, from
Figure~\ref{fig: GrowthRatio}, we notice that the growth ratio for
cascades hits a peak around size 32, and is the same for the smallest
and largest cascades!

Thus, it seems clear that even though large cascades (especially
``viral'' ones) are infrequent on Twitter, they constitute a
substantial fraction of audience attention. This observation leads to
the question of how does the audience react to the presence of
cascades in their home timeline. We address this question in the next
section by analyzing user engagement.

\subsection{Engagement with Cascades}
Users on Twitter engage with tweets in a variety of ways, and we focus
on the following engagements in our analysis: retweets (resharing the
tweet with your followers), likes (previously known as favoriting),
and clicks (either a click on a link/mention/hashtag in a tweet, or a
visit to a ``tweet details'' page are considered as clicks). Together,
these engagements provide a broad perspective on how users perceive
the content as engagements are often a reflection of how much users
enjoyed the content. This is not always the case though, and
engagement is skewed by a range of factors: social acceptability,
context, and inherent clickability (for instance, ``clickbait'' may
have a high clickthrough rate) of content to name a few. Despite these
shortcomings, the large scale of data analysis that we conduct does
provide a directional guide on user enjoyment by measuring engagement.

\begin{figure}
    \centering
    \includegraphics[width=0.8\linewidth]{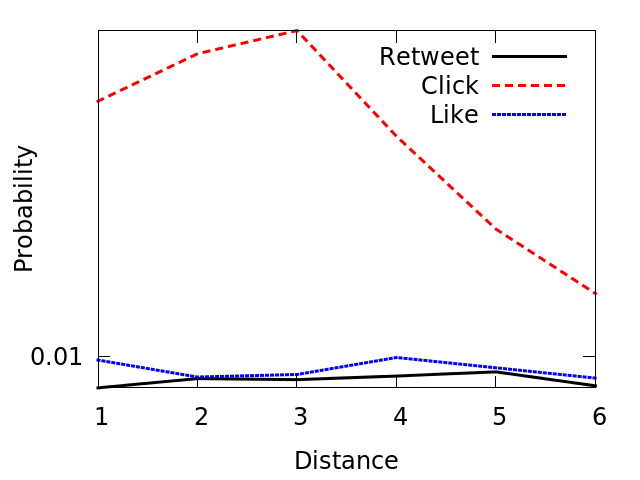}
    \caption{Probability of interaction based on distance. Note that
      the $y$-axis values are randomly pinned to 0.01.}
    \label{fig: InteractionProbVsDistance}
\end{figure}

Engagement with cascades doesn't occur in a vacuum, and here we
contrast engagement on cascades against the natural baseline of
engagement on ``organic'' tweets, i.e. tweets authored directly by a
user's followings. This comparison is readily obtained by measuring
how the probability of retweets, likes and clicks varies with graph
distance. These curves are presented in Figure~\ref{fig:
  InteractionProbVsDistance}, which shows that the various engagement
measures behave quite differently. In particular, a tweet from a
neighbor has a higher probability of receiving a like than tweets
coming from users that are farther away in the network. On the other
hand, a cascade tweet that originated outside of the user's direct
network has a higher chance of getting retweeted and clicked. This
perhaps is an indication that liking has a social element to it, and
users tend to primarily like personalized content or tweets that come
from their direct neighbors. On the other hand, a tweet that arrives
in a cascade from outside the neighborhood is ``retweetable'' by
definition, and hence just by this selection mechanism it increases
its chances of getting retweeted as compared to an average tweet that
may not received any retweets at all. The click probability curve
shows different behavior than both likes and retweets, and the
probability of clicking on the tweet {\em increases} till distance
3. It is important to point out that to a consumer on Twitter, the
only visible distinction in distance is whether it is 1 (direct
connection) or greater (retweet). Thus, the difference between click
probability in distances 2 and beyond likely comes from something
other than user selectiveness between in and out of network
content. An appealing hypothesis is that perhaps inherently clickable
content travels farther on the network. We examine this next via a hop
count analysis.

\begin{figure}
  \centering
  \includegraphics[width=0.8\linewidth]{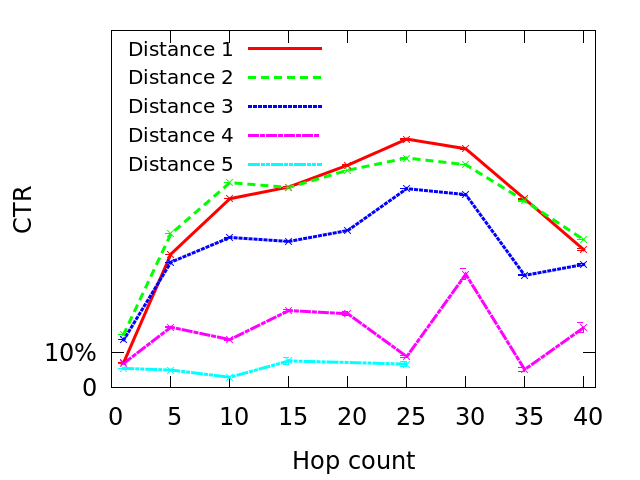}
  \caption{The click Through Rate (CTR) of tweets coming from
    different distances against different hop counts with
    errorbars. Note that the $y$-axis values are randomly pinned to
    10\%.}
  \label{fig: Geodesic-HopCount}
\end{figure}

Recall that hop-count refers to the distance from the user to the
author in the cascade tree. In order to understand the click
probability variation, we turn to examining click-through rate (CTR)
of tweets by hop-count --- since this route is predominantly available
to larger cascades, it provides us a way to measure how users react to
large cascades versus other smaller cascades. This data is presented in
Figure~\ref{fig: Geodesic-HopCount}, where there is a curve for a
fixed distane showing the average CTR of tweets based on the hop-count
value. There are several things to notice in this plot. First, observe
that for a fixed hop-count, CTR generally decreases with distance ---
this clarifies that the increase with distance observed in
Figure~\ref{fig: InteractionProbVsDistance} is at least partially due
to an effect that it to some extent like Simpson's paradox.
We also note from Figure~\ref{fig:
  Geodesic-HopCount} that CTR generally increases with
hop-count. Recall that higher hop counts are generally only available
to large cascades, and hence the higher CTR indicates that popular
content generates more clicks. We emphasize that this is not a causal
statement, and in particular popular content might precisely be more
popular {\em because} it generates more clicks.

\begin{figure}[hbtp]
\begin{subfigure}[t]{.23\textwidth}
\centering
\includegraphics[width=\linewidth]{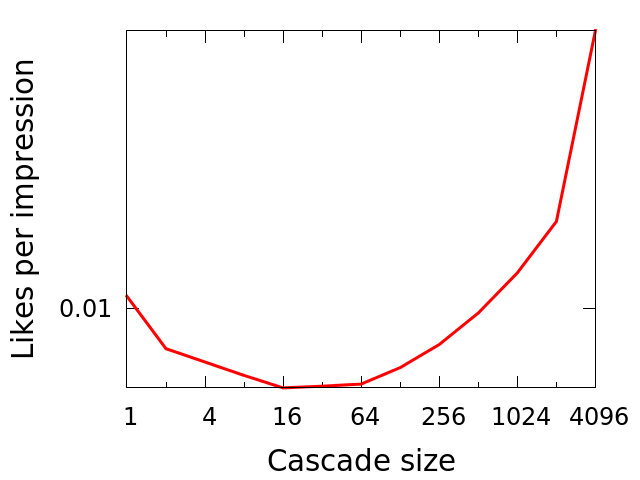}
 \caption{}\label{fig: U-ShapedCurve}
\end{subfigure}
\begin{subfigure}[t]{.23\textwidth}
\centering
\includegraphics[width=\linewidth]{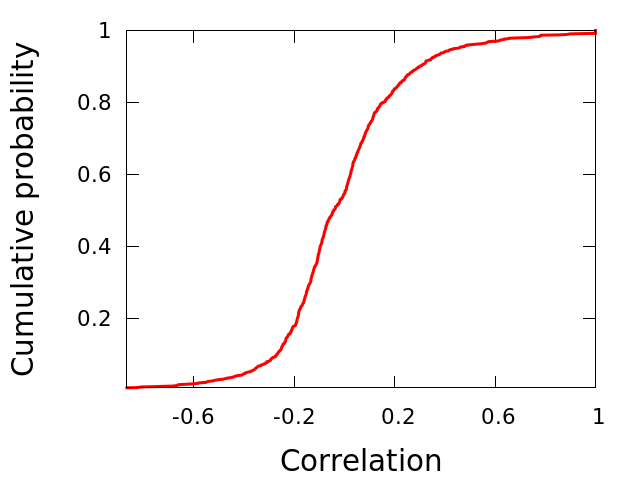}
\caption{}\label{fig:SizeLikeCorrelation}
\end{subfigure}

\caption{(a) Likes per impression for different cascade sizes bucketed
  by log base 2 (note that the $y$-axis values are randomly pinned to 0.01.) (b)
Distribution over the correlation of different users, liking content based on its size}
\end{figure}
%
%

Given this data, we are left with the intriguing observation that
users enjoy consuming (via liking/clicking) content both from their
direct neighbors as well as from large cascades. But the combined
effect of these two mechanisms is apriori unclear. In particular,
since large cascades can travel farther from their source, does their
overall appeal overcome the fact that their audience is not their
direct neighbors? The most direct way to study this question is to
look at like probability per tweet impression based on the cascade
size of the tweet, and Figure~\ref{fig: U-ShapedCurve} shows the
result of this investigation. Note that the cascades are bucketed as
before, since the number of cascades with large sizes are rare. We
observe from Figure~\ref{fig: U-ShapedCurve} that cascades with very
small and very large sizes have high like per impression rates. These
stand in contrast with a smaller like per impression value for medium
size cascades. We can thus see the two mechanisms mentioned above at
play here: at small cascade sizes, content is being liked by neighbors
(perhaps driven by a social aspect), while at large sizes content is
likeable in general so most users enjoy consuming it. But there is an
``uncanny valley'' in the middle where not naturally likeable content
reaches users who are not quite interested in it. The connection between
Figure~\ref{fig: GrowthRatio} and \ref{fig: U-ShapedCurve}
is very interesting, and we leave further investigation of this
connection as a direction for future work.

\begin{figure*}
    \begin{minipage}[l]{\columnwidth}
        \centering
        \includegraphics[width=0.91\linewidth]{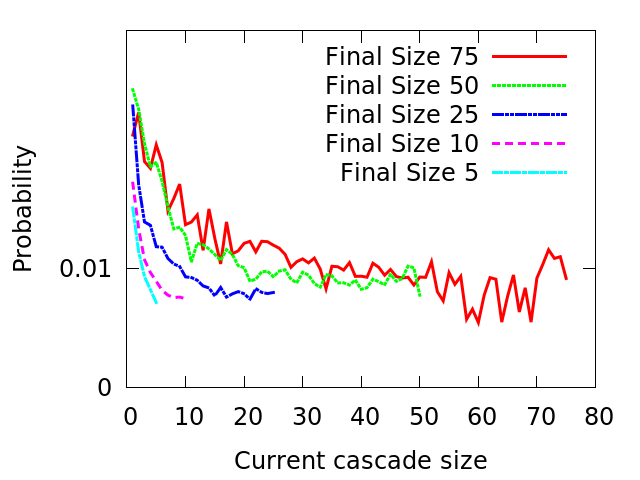}
        \captionsetup{labelformat=empty}
        \caption{(a) Likes}\label{fig: CurrentCascadeSize_A}
    \end{minipage}
    \begin{minipage}[r]{\columnwidth}
        \centering
        \includegraphics[width=0.9\linewidth]{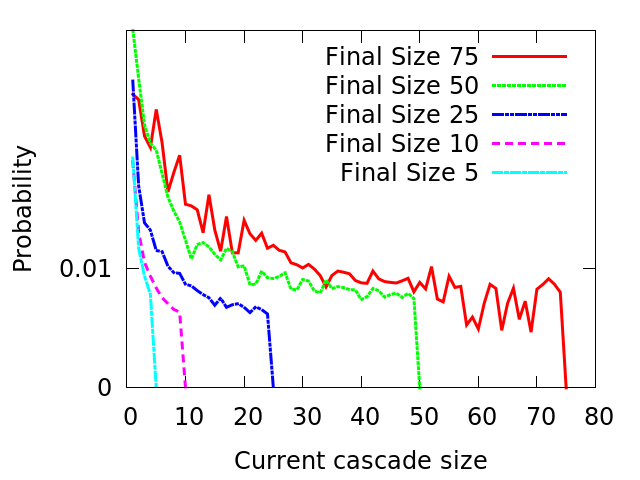}
        \captionsetup{labelformat=empty}
        \caption{(b) Retweets}\label{fig: CurrentCascadeSize_B}
    \end{minipage}
    \addtocounter{figure}{-1}
    \addtocounter{figure}{-1}
\caption{The behavior of consumers relative to the current popularity of the content}
 \label{fig: CurrentCascadeSize}
\end{figure*}

From \ref{fig: U-ShapedCurve}, it seems clear that the most popular
content, as indicated by size, has the highest like rate. Since users
on Twitter can see the overall popularity of a tweet (the number of
retweets and likes), size does provide a signaling mechanism that
could potentially bias the highest like rate in the favor of popular
tweets, fueling a rich-gets-richer effect. Another possibility is that
tweets have an intrinsic ``quality'' that drives their popularity and
higher size is partially a result of this quality (it could still
involve other factors too, such as being lucky and receiving attention
from popular users early in the tweet's lifetime). To disambiguate
between these possibilities, we observe that over the lifetime of a
cascade, different users view the tweet at different points in its
popularity. This allows us to study whether users react differently to
tweets that had different eventual sizes but were viewed at the same
level of popularity by users. We can see from results in
Figure~\ref{fig: CurrentCascadeSize} that the cascades that ended up
with a larger eventual size had a higher like and retweet rate even
earlier in their life. This provides some evidence that intrinsic
quality of a tweet does contribute to its eventual popularity.

\subsubsection{Individual User Preferences}
The analysis above suggests that globally popular content is also
locally popular at an aggregate level for users. We now examine
individual user variation for these preferences: do most users like
globally popular content? Are users consistent in their preferences on
local vs global content? In order to study these questions, we
randomly selected $10000$ {\em active} users on Twitter, where if a user
had at least one interaction each day for more than $15$ days out of a
$16$ day period in June, we count him/her as an active user.

Let us first turn to the question of local vs global preference for a
user.  To study this, for each user we compute the Pearson correlation
coefficient between the final cascade size and like probability of all
the tweets that the user viewed in this period. The distribution of
these correlations over the $10000$ users is shown in
Figure~\ref{fig:SizeLikeCorrelation}. As we see from the plot, most
users have a negative correlation, implying that they prefer
personalized content. However, the correlation coefficient is low for
these negatively correlated users, and if we only focus instead on
users who have strong correlation then most of these are positively
correlated. In either case, users do seem to exhibit a
local/global content preference.

The local/global preference as expressed by a correlation is however
rather weak, and leaves open the question of whether users are
consistent in their preferences. Here, we study user consistency via
the following analysis. We define the function $f(u,d,x)$, for user
$u$, day $d$ and any $x \in [0,1]$, to be the fraction of likes
produced by the $x$ fraction of
smallest cascades that user $u$ sees on day $d$.
If $f(u,d,0.5) < 0.50$, it means that user
$u$ likes content that is part of smaller cascades at a greater
rate than they like content that is part of larger cascades;
we could think of this as the user liking personal content
more than broadly-shared general content.
The implication is the opposite if the inequality is reversed:
if $f(u,d,0.5) > 0.50$, then the user $u$ like content that it
part of large cascades (and hence more broadly-shared general
content) at a greater rate.
We operationalize this definition by identifying users who like small
(respectively, large) cascades according to the condition
$f(u,d,0.5) \geq 0.55$ (respectively $f(u,d,0.5) \leq 0.45$).
\begin{table}
\centering
\caption{Fraction of each type of the three user.}
\begin{tabular}{|c|c|}
\hline
Consistent small cascade liker & 19.4\% \\ \hline
Consistent big cascade liker &  47.1\% \\ \hline
Indifferent & 33.5\%\\ \hline
\end{tabular}
\label{table: UserConsistency}
\end{table}

Further, if a user likes the same type of cascades at least $11$ days
over the $16$ day period, we call her a consistent user (either a
liker or large cascades or a liker of small cascades); otherwise we call her
\textit{indifferent}. With this definition of consistency, we find
that more than $66\%$ of our users are consistent.  The fraction of
users of each type can be seen in Table \ref{table: UserConsistency}.
As a simple baseline, note that if each user independently decided
each day with uniform probability whether to like small cascades
or large cascades, we'd expect only $21\%$ of all users to be consistent,
rather than over 66\% as we find in the data.


\begin{figure*}[hbtp]
\centering
\begin{subfigure}[t]{.31\textwidth}
\centering
\includegraphics[width=\linewidth]{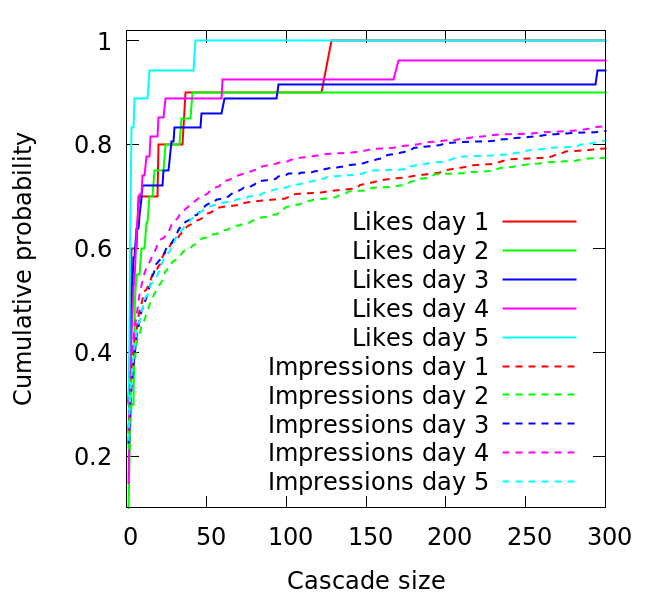}
 \caption{Small cascade liker}\label{fig: UserConsistency_a}
\end{subfigure}
\begin{subfigure}[t]{.31\textwidth}
\centering
\includegraphics[width=\linewidth]{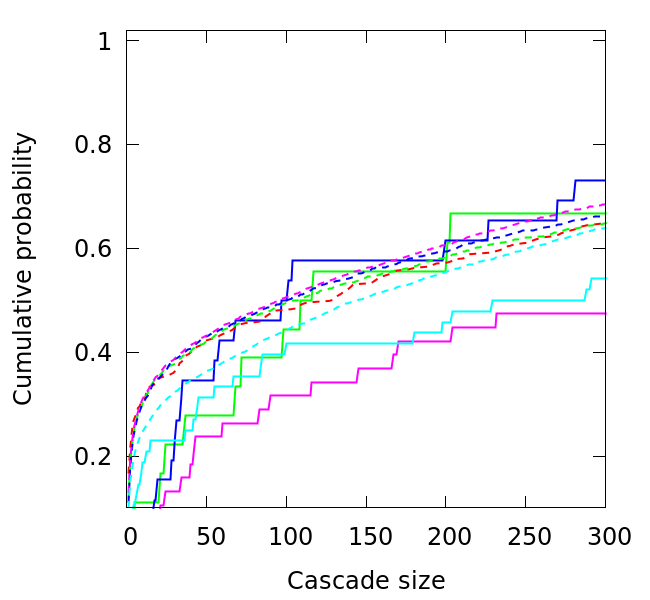}
\caption{Large cascade liker}\label{fig: UserConsistency_b}
\end{subfigure}
\begin{subfigure}[t]{0.31\textwidth}
\centering
\includegraphics[width=\linewidth]{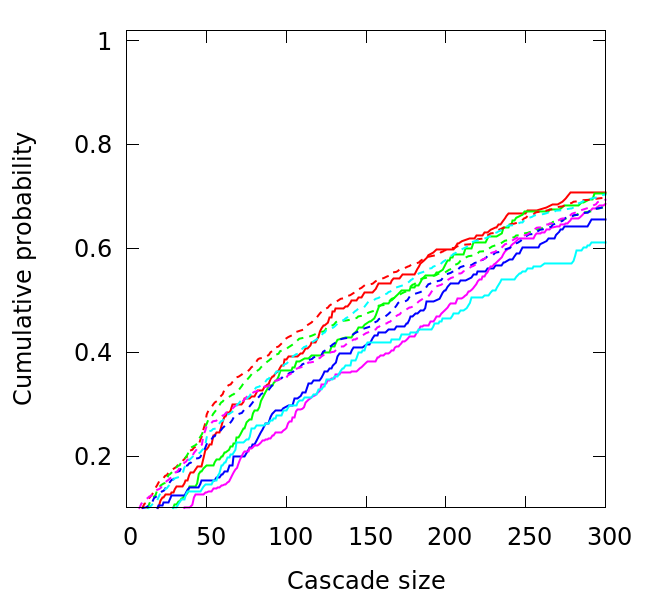}
\caption{Indifferent}\label{fig: UserConsistency_c}
\end{subfigure}
\caption{Three samples of different groups of users. Figure (b) and (c) have the same legends as figure (a).}
\label{fig: UserConsistency}
\end{figure*}

We also provide a useful illustration of how these three user
behaviors are distinct. We define $i(u,d,x)$ ($l(u,d,x)$) to be the
fraction of impressions (likes) that user $u$ had on day $d$ and the
cascade sizes were less than $x$. We then show how these two functions
behave for all three types of users (prefers personal content, broader
content or indifferent) in Figure~\ref{fig: UserConsistency}. We draw
the function $i(u,d,x)$ with dashed line and $l(u,d,x)$ with solid
lines. This data clearly indicates that users do indeed have
consistently different preferences, which might stem from using
Twitter for different purposes. We leave a more in-depth investigation
of this to future work, but note that Twitter does exhibit both social
and information network structural properties indicating the presence
of multiple usage scenarios~\cite{MSGL14}.

We now summarize our overall empirical findings: cascades have a large
audience on Twitter, larger than one might expect just based on
cascade occurrence (cf. the {\em Impressions Paradox}). Furthermore,
users seem to like these cascades, even relative to organic
content. However, these observations seem to be somewhat at odds with
each other: since a lot of users see large cascades, either cascading
content is always appreciated by a large majority or somehow it
reaches predominantly users who would enjoy it. The latter possibility
is what we explore in the remainder of this work, with the lens of a
simple theoretical model.

\section{Modeling Cascades For Audience}
The empirical analysis presented earlier leads to the conclusion that
a quarter of a typical user's timeline consists of retweets, and
users find this content engaging. As we've alluded to before, this
presents a conflict between the view of users choosing exactly what
content they want to see (by following a set of users) and engaging
content being bubbled up through the network via retweets. In this
section, we posit a simple theoretical model that shows this
conflict can be resolved in a natural manner by users being selective
about content they retweet. We emphasize that the model is purposely
bare-bones so that it can provide stylized insight
into why cascades turn out to be relevant for users.

The main idea of our model is to think of an {\em inverted} tree where
a member of the audience (a consumer of cascades) is the root,
in contrast to the standard view of the cascade
originator as the root node. This allows us to examine the path
that content takes to arrive at the audience member at
the root node.  The model includes a notion of
{\em topic}, which governs whether a given user is
interested in a given tweet. We can now define the model formally
using these notions.

Each user $u$ in the network has a set of topics they are interested
in: $I_u \subset I$, where $|I_u|=d$ and the set $I$ contains the
universe of topics ($|I| = D$). For a fixed arbitrary user $a$, we
consider their two-hop neighborhood
(recall from Table \ref{table: ImpressionsOnDistanceAndHopCount} that
vast majority of the retweets arrive
from two hops away): $a$ follows the set of
users $B=\{b_1,\ldots,b_k\}$, and each user $b_i$ follows
$C_i=\{c_{i1},\ldots,c_{ik'}\}$, with the entire second hop
neighborhood being denoted by $C=\bigcup\limits_{i=1}^k C_i$. In this
simple network, we assume that if a user $u$ follows user $v$, then
$u$ is interested in a significant fraction of $v$'s topics. More
formally, we assume that for a fixed constant $0<\alpha<1$,
$|I_u \cap I_v| \geq \alpha d$. We'll shortly address the question of
how this extended neighborhood is generated, but first we will focus
on the cascade process given such a network.

Given a network that has content being produced and consumed on it,
for each user $u$ we will have a ``home timeline'' which is populated
by content produced by the set of users that $u$ follows; we denote the set of
tweets in $u$'s home timeline by $\TL{u}$. We now specify the content
production process. First, we link users' topical interests to tweets
by assigning a single topic $i_t$ to each tweet $t$, where $i_t$ is
selected from the author's list of topics. We also assume that
tweet $t$ has an intrinsic {\em quality} $q_t$, which is in line with
our observations from Figure~\ref{fig: CurrentCascadeSize}.
For tweet production, we assume simplistically that at
all users produce tweets at each epoch as follows. First, each user
$u$ creates an original candidate tweet $t_o$ on one of the topics
$i_{t_o}\in I_u$, with a quality $q_{t_o}$ drawn from a specified
distribution $\mathcal{D}$, but doesn't publish this candidate to its
followers yet.
The user picks her tweet that she will publish as follows:
she considers her own tweet and her
home timeline $\TL{u}$  --- consisting of tweets produced in the previous
epoch by the users she follows --- and then among the tweets $t$
in this set for which $i_{t} \in I_u$, she selects the
tweet $t_h$ of highest quality $q_{t_h}$

This setting reflects the fact that users can both participate in
cascades as well as produce original content. Further,
the model allows for user curation for cascades that biases towards
participation in higher quality cascades. The goal of the model is
to capture the consumer viewpoint on cascades, and the
consumer view is governed by whether the content in their home
timeline is high quality, and also on whether it is on a topic that is
interesting to them. We formally define these metrics for the home
timeline of a given user $a$ as follows:
\begin{definition}[Precision]
  We define {\em precision} for a user $u$ as the fraction of tweets in
  $u$'s timeline that $u$ would be interested in:
  $\precision{u} = \sum_{t \in \TL{u}} \frac{|\{i_t \in I_u\}|}{|\TL{u}|}$.
\end{definition}
\begin{definition}[Quality]
  We define the timeline {\em quality} for a user $u$ as the average quality
  of all tweets in $u$'s home timeline:
  $quality(u) = \sum_{t \in \TL{u}} \frac{q(t)}{|\TL{u}|}$.
\end{definition}
\begin{definition}[Timeline Utility (TLU)]
  If $\delta$ is the quality coefficient for tweets that are not on topic for
  the consumer then the overall {\em timeline utility} of the home timeline for a given
  user $u$, can be defined as
  $$\TLU{u} = \sum_{t \in \TL{u}} g(t,u) \cdot q_t.$$
  Here $g(t,u) = 1$ if $i_t\in I_u$, and $g(t,u) = \delta \leq 1$ otherwise.
  Thus,
  $\TLU{u}$ increases if there are high-quality tweets and decreases
  if there are off-topic tweets.
\end{definition}

Thus, the model aims to capture the effect of cascades on the twin
goals of having users enjoy both high quality and precise
content. Intuitively, retweets seem to filter for high quality tweets
but the effect on precision is less clear. In fact, the exact effect
on precision depends on network topology. We now define how the
network is created, starting with a simple model that is
analytically tractable. We'll later define a more complex model on
which we simulated the model.

Possibly the simplest way to construct a network is to build a
tree. We proceed by having given the node $a$ for which we want to study
the above metrics, and her corresponding topical interests. Then we
pick $k$ random interest sets that satisfies the homophily condition
(of $\alpha$ fraction interest overlap), and designate those to be the
interest sets of the nodes in $B$. We then repeat the same procedure
for each $b_i$ and create $k$ neighbors for each $b_i$ while also
satisfying the homophily restrictions. Given this network generation
model, now we can analyze the effect of cascades on previously defined
quality and precision metrics.

For the theoretical analysis we look into a basic setting where
$\delta=1$. In the next
subsection we remove these restrictions and report the results on the simulation.
We begin by noting that based on the graph generation process, the
distribution of topics in $C_i$ on the topics in $I_{b_i}$ is
uniform.
Now, in every epoch but the first (once retweets start moving through
the network), the view of a node $b_i$ is as follows.
If $b_i$ receives a tweet that she is not interested in, that has no
chance of being propagated. Otherwise, she will keep the tweet as a
candidate in the maximization step. Now, we know that the distribution
of the tweet topics coming from $\{c_{i1},\ldots,c_{ik}\}$ is uniform
on $I_{b_i}$, so we observe that the model is equivalent to $b_i$
itself generating many tweets and only publishing the best. If we look
at the process in this manner, then it is clear that the expected
precision does not change when we introduce retweets. However, with
this procedure the quality of tweets will go up. It is easy to see
that the expected quality of a single tweet is less than the
expectation of the maximum of independent draws from the same
distribution. We formalize the gain in quality for two specific
distributions, and note that the analysis can be extended to other
distributions.

In particular we investigate the two cases where $\mathcal{D}$ is
either a uniform distribution over $[0,1]$ or an exponential
distribution with rate $\lambda$.  We use $\mathcal{G}$
to refer to the distribution of the
maximum of $k$ draws from $\mathcal{D}$.
A question we want to answer is; how much do retweets help the quality
of the timeline to increase?  Let $X$ be a random variable drawn from
$\mathcal{D}$ and $Y$ be a random variable drawn from $\mathcal{G}$.
With our notation we are interested in
$\frac{\mathbb{E}[Y]}{\mathbb{E}[X]}$.  It is well known that the mean
of a uniform distribution over $[0,1]$ is $\frac{1}{2}$ and the mean
of a exponential distribution of rate $\lambda$ is
$\frac{1}{\lambda}$.  Now we state two well-known lemmas about the maximum of
a set of independent draws from a fixed distribution \cite{Ross2006}.
\begin{lemma}
  The expectation of the maximum of $k$ i.i.d draws from a uniform
  distribution over $[0,1]$ is $\frac{k}{k+1}$.
\end{lemma}
\begin{lemma}
  The expectation of the maximum of $k$ i.i.d draws from an exponential
  distribution of rate $\lambda$ is $\frac{H_k}{\lambda}$,
  where $H_k$ is the $k$-th harmonic number.
\end{lemma}
We can now state the main theoretical result on the effect of retweets
on quality. We skip by proof but note that it is easily obtained from
the above two lemmas.
\begin{theorem}
  The multiplicative increase in quality
  $\frac{\mathbb{E}[Y]}{\mathbb{E}[X]}$ in the scenario where $\mathcal{D}$ is
  uniform is $\frac{2k}{k+1}$ and when $\mathcal{D}$ is an exponential
  distribution with rate $\lambda$, it is $H_k$.
\end{theorem}

The theorem formalizes the gain in quality that comes from having
cascades in the network, which is a counterfactual that is not easily
observable in the Twitter network\footnote{This remains hard to
  measure even via experimentation as there is a large network effect
  to contend with.}. To gain some sense of the scale of this increase,
let us consider a network where the average degree of nodes is
$50$ (This number is a lower-bound for the Twitter network).
By plugging in $50$ for $k$ we see that the uniform and
exponential distribution model yield a $96\%$ and $350\%$ increase in
quality, respectively! This shows how valuable retweets can be to a
network, as illustrated by our stylized model.
We re-emphasize that
we do not claim our theorem to be representative of the gain in the
Twitter setting --- the goal of this modeling exercise is to shed
light on the effect of cascades on the audience, and we can quantify
that effect with the help of a formal model. We now explore
generalizing this model by removing some assumptions.

\subsection{Model Extensions}
The above model is quite simple and makes a few strong assumptions in
order to be analytically tractable. In this section, we explore the
effect of removing or generalizing these assumptions via simulating
the model. In particular, there are two obvious ways we can generalize
the model.

First, we previously assumed that the two-hop graph was a tree, but
now we generalize that to the following two graphs:
\squishlist

\item \textbf{Tree Contracted model:} Here, we create the graph using
  the Tree model but contract all the nodes in layer $B$ and $C$ that
  have the same interest set into one single node.
\item \textbf{$k$-NN: } Here, we create a number of nodes\footnote{For
    our simulations, we use $4 \times 10^5$ nodes.} with their own
  interest set, and each person follows the $k$ people that have the
  most common interests with them.  (We break ties in this choice of
  $k$ at random.)

\squishend

Second, we can change various aspects of the model in the previous
section, which includes both changing existing parameter values and
generalization of behavior from the model in the previous
section. The generalizations we examine are as follows:
\squishlist
\item \textbf{$\delta$:} Recall that if a user receives an off-topic
  tweet of quality $q$, then she sees that as a tweet with quality
  $q\delta$. This parameter is used for both in finding the $TLU$ and
  the retweeting procedure, and was previously set to 1. Here, we
  explore the effect of using other values for $\delta$.
\item $k'=\frac{|C|}{|B|}$: This parameter allows us to control the
  average degree of the neighbors of the target node to her
  degree. We note that in our simulations, the results stabilize once
  this ratio is above $20$, and this value is above 20 for Twitter in
  particular.
\item \textbf{The self-interest factor (p)}: We introduce a new
  parameter to allow users the flexibility to tweet original content
  instead of simply retweeting others. We stipulate that in each
  epoch, a user tweets her own tweet with probability $p$, and
  otherwise, retweets one of the tweets that she has received. Also,
  when retweeting the user differentiates between the tweets from the
  people she follows and the people that she does not follow (We take
  following to be a proxy for knowing).  Thus, once she decides to
  retweet someone, with probability $p$ she picks the highest quality
  tweet created by one of her immediate followees, otherwise she picks
  the highest quality tweet from the pool of tweets coming from more
  than one hop away from her. \\
  Note that as we increase $p$, this method creates a bias towards (i)
  creating her own organic tweet, and (ii) while retweeting, she
  prioritizes her immediate neighbors' tweet over tweets coming from
  deeper in the network. This creates a mechanism that constructs
  timelines which most of its content is from at most two hops away.
  \squishend

\begin{figure}
\centering

\begin{subfigure}[t]{.23\textwidth}
\centering
\includegraphics[width=\linewidth]{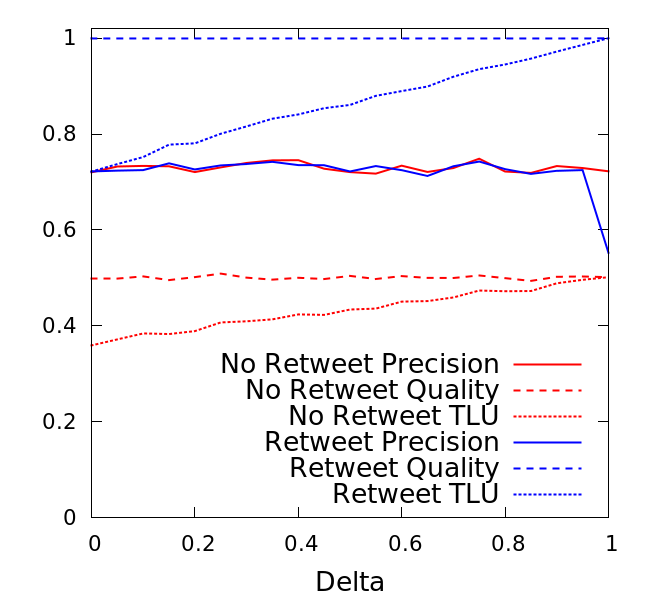}
        \caption{}\label{fig: Simulation_a}
\end{subfigure}
\begin{subfigure}[t]{.23\textwidth}
\centering
\includegraphics[width=\linewidth]{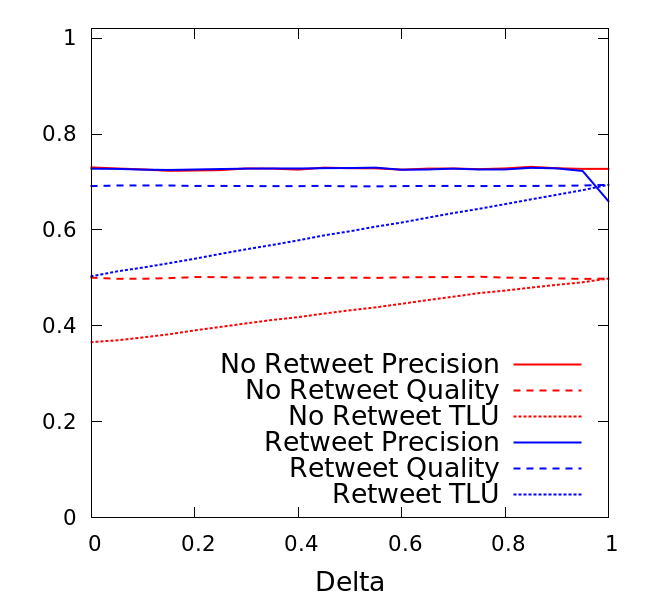}
\caption{}\label{fig: Simulation_b}
\end{subfigure}

\medskip

\begin{subfigure}[t]{.23\textwidth}
\centering
\vspace{0pt}
\includegraphics[width=\linewidth]{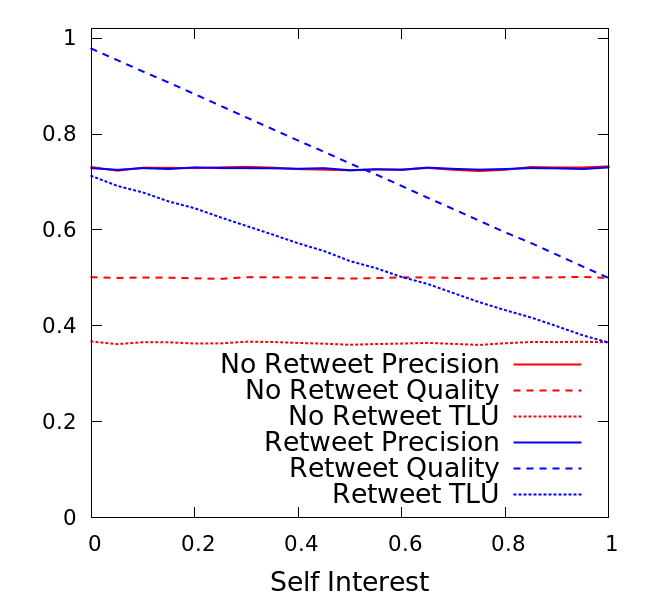}
\caption{}\label{fig: Simulation_c}
\end{subfigure}
\begin{subfigure}[t]{.23\textwidth}
\centering
\vspace{0pt}
\includegraphics[width=\linewidth]{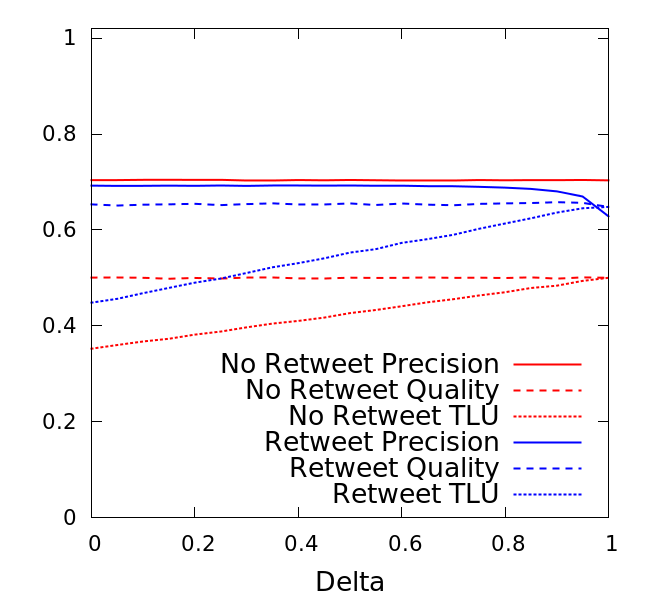}
\caption{}\label{fig: Simulation_d}
\end{subfigure}

\begin{minipage}[t]{.5\textwidth}
\caption{Precision, Quality and TLU over different models and parameters set. (\subref{fig: Simulation_a}) Simple tree model (\subref{fig: Simulation_b}) Tree Contracted model with p = 0.6 (\subref{fig: Simulation_c}) Tree Contracted model with $\delta =0$ (\subref{fig: Simulation_d}) $k$-NN model with p=0.6}
\label{fig:Simulations}

\end{minipage}
\end{figure}

The simulations results from varying the network model and the above
parameters are all shown in Figure \ref{fig:Simulations}. From the
results, we observe that the following holds unless $\delta$ is close
to $1$\footnote{Recall that $\delta$ being close to 1 would imply
  users not distinguishing between on and off topic tweets, and hence
  it makes sense to focus on results where $\delta$ is smaller than
  1.}: by introducing retweets in the network, the precision remains
essentially the same and the quality goes up, leading to a higher
$TLU$. This is quite consistent with the theorem in the previous
section and shows that the observations in the previous section are
somewhat robust to the specific network model and parameters.

\section{Conclusions}

Information cascades on networks affect not only the users who
actively participate in them, but also the audience who are
consequently exposed to the cascades. Our work provides a view of
cascades from the point of view of the audience, which has not
received much attention to the best of our knowledge. Our findings
related to the {\em Impressions Paradox} provides a novel perspective
for future work on the effect of cascades on their audience.

The theoretical model presented here is quite simplistic, but it does
highlight the crucial role of cascade participants as gatekeepers of
precision. In addition to further generalizations of the model,
this work raises several additional open questions for future work.
Clearly, not all
users retweet on topic, but does the network rewire itself (via
audience following/unfollowing) so that precision remains high?
Furthermore, an aspect of cascades we did not discuss here is their
usefulness as a discovery mechanism; can effective discovery coexist
with high precision in the network?

Healthy dynamics on social networks requires both active producers
and engaged consumers. We believe our work provides a novel and useful
consumer counterpoint to the extensive literature on the role that
producers play in cascades.
\bibliographystyle{abbrv}

\end{document}